\documentclass[aps,showpacs,showkeys,superscriptaddress]{revtex4}

\usepackage{graphicx}

\begin{document}
\title{Electronic properties of disclinated nanostructured cylinder}

\author{R. Pincak}\email{pincak@saske.sk}
\affiliation{Bogoliubov Laboratory of Theoretical Physics, Joint
Institute for Nuclear Research, 141980 Dubna, Moscow region, Russia}
\affiliation{Institute of Experimental Physics, Slovak Academy of Sciences,
Watsonova 47,043 53 Kosice, Slovak Republic}

\author{J. Smotlacha}\email{smota@centrum.cz}
\affiliation{Faculty of Nuclear Sciences and Physical Engineering, Czech Technical University, Brehova 7, 110 00 Prague,
Czech Republic}

\author{M. Pudlak}\email{pudlak@saske.sk}
\affiliation{Institute of Experimental Physics, Slovak Academy of
Sciences, Watsonova 47,043 53 Kosice, Slovak Republic}

\date{\today}



\def\wu{\widetilde{u}}
\def\wv{\widetilde{v}}

\begin{abstract}
The electronic structure of nanocylinder without and with a small
perturbation is investigated with the help of calculation of the
local density of states. A continuum gauge field-theory model is
used for this purpose. In this model, Dirac equation is solved on a
curved surface. The local density of states is calculated from its
solution. The case of 2 heptagonal defects is considered. This paper
is an extension of our previous work \cite{1} where one heptagonal
and one pentagonal defects in hexagonal graphene network were
compared. The metallization for the perturbed cylinder structure is
found.
\end{abstract}

\maketitle

\section{Introduction}\

The carbon nanostructures play a key role in constructing nanoscale devices like quantum wires, nonlinear
electronic elements, transistors, molecular memory devices or electron field emitters. Their molecules are variously-shaped geometrical forms its surface is composed of disclinated hexagonal carbon lattice. The main structure of this kind is graphene - the carbon lattice plane from which all other kinds are derived.\\

The most famous form is fullerene - a material composed of molecules which have the form of a soccer ball \cite{2}. Other kinds are nanocones, nanotubes, nanotoroids, nanocylinders, nanoribbons etc. The various forms of the nanostructures are ensured by the topological defects in the graphene which are most often presented by the pentagons and heptagons in the hexagonal plane lattice. In the closest vicinity of these defects, the positive resp. the negative curvature arises for the case of the pentagons or the heptagons, respectively. Generally, for the $n$-sided polygon, $n<6$ corresponds to the positive curvature and $n>6$ corresponds to the negative curvature. This fact coincides with the choice of the form of the defects in the particular cases. There is no heptagonal defect in the fullerene, but a lot of these kinds of defects can be found in many open forms of nanostructures. But we can find them also in some of the closed forms like nanotoroids or more complicated, folded forms of nanotubes. Most often, the heptagons appear in pairs with pentagons in the connecting parts of the folded forms \cite{4}.\\

Because of the applications, the research of the electronic properties of the carbon nanostructures is important. One of the main characteristics is the local density of states ($LDoS$). In the presented model coming from the effective-mass theory, knowledge of the solution of the corresponding Dirac equation is necessary for the calculation \cite{mele}. This solution is represented by the wave-function and to find it, we have to know the geometry of the molecular surface.  From the mentioned facts follows that the chosen geometry can be only an approximation of the complicated real situation: for example, the spherical geometry of fullerene is not suitable for the description of the closest vicinity of the defects but it correctly describes the properties of the whole molecule. As discovered in \cite{5} for the case of nanocones, the most suitable geometry for the description of the close vicinity of the defects is the hyperboloidal geometry. Very often, for a given geometry, the number of possible defects is limited.\\

The solutions for spherical, conical and 2-fold-hyperboloidal cases were found in \cite{3,6,5}. In \cite{1}, we used the presented model for calculation of the electronic properties of the structures with the geometry of the 1-fold hyperboloid. The aim was to describe the electronic properties in the vicinity of the locally negative curvature of an arbitrary nanoparticle. This restriction did not enable us to do the calculations for the case of more than 1 defect. In this paper, we present a model describing the electronic properties of a simple nanocylinder and a curved nanocylinder including 2 heptagons at the opposite sides of the surface. The hyperboloidal geometry is used again. Because a nanocylinder is an opened nanotube, comparison with the case of the  capped nanotube could be performed (see e.g. \cite{vanhove} for this purpose).\\

This paper is organized as follows: the second section describes the computational formalism. The third section summarizes the basic properties of nanotubes and derived nanostructures. The fourth section researches electronic properties of cylinder without and with a defect.  In the fifth section, a small review about the graphene nanoribbons and their properties is given and calculation of its $LDoS$ is performed. In Conclusion, the obtained results for cases of cylinder and inifinitely long nanoribbon are compared and discussed and a brief review about the production of the nanocylinders is introduced. For the cylinder, the normalization constants are computed in the Appendix A and zero modes for defect-free and perturbed case are computed in the Appendices B and C, respectively.\\\\

\section{Computational formalism}\

To research the electronic properties, we have to solve the Dirac equation in (2+1) dimensions. It has
the form
\begin{equation}\label{DirC}i\sigma^{\alpha}e_{\alpha}^{\mu}[\partial_{\mu}+\Omega_{\mu}-ia_{\mu}-ia_{\mu}^W]\psi=
E\psi,\end{equation}
where $\partial_{\mu}$ means the partial derivation according to the $\mu$ parameter, i.e. $\partial_{\mu}=\frac{\partial}{\partial x^{\mu}}$.\\

In this equation, besides the energy $E$, the particular constituents have the following sense:\\\\
$\sigma^{\alpha}, \alpha=1,2,$ denote the Pauli matrices.\\\\
The zweibein $e_{\alpha}^{\mu}, \mu=z,\varphi$ stands for incorporating fermions on the
curved 2D surface and it has to yield the same values of observed
quantities for different choices related by the local SO(2)
rotations:
\begin{equation}e_{\alpha}\rightarrow e_{\alpha}'=\Lambda^{\beta}_{\alpha}e_{\beta},\hspace{1cm}
\Lambda^{\beta}_{\alpha}\in SO(2).\end{equation} For this purpose,
 a covariantly-constant local gauge field $\omega_{\mu}$ is incorporated \cite{7}:
\begin{equation}\partial_{\mu}e^{\alpha}_{\nu}-\Gamma^{\rho}_{\mu\nu}e^{\alpha}_{\rho}+
(\omega_{\mu})^{\alpha}_{\beta} e^{\beta}_{\nu}=0,\end{equation}
where
\begin{equation}
\Gamma^{\rho}_{\mu \nu}=\frac{1}{2}g^{\rho\tau}\left(\frac{\partial g_{\tau\nu}}{\partial x^{\mu}}+\frac{\partial g_{\mu \tau}}{\partial x^{\nu}}-
\frac{\partial g_{\mu \nu}}{\partial x^{\tau}}\right)
\end{equation}
is the Levi-Civita connection coming from the
metrics $g_{\mu\nu}$ (see below). Then $\omega_{\mu}$ is called the
spin connection.\\\\
The constituent
\begin{equation}\Omega_{\mu}=\frac{1}{8}\omega^{\alpha\beta}_{\mu}[\sigma_{\alpha},\sigma_{\beta}]
\end{equation}
denotes the spin connection in the spinor representation. Its components are
\begin{equation}\Omega_z=0,\hspace{1cm}\Omega_{\varphi}=i\omega\sigma_3,\end{equation}
where
\begin{equation}\omega=\frac{1}{2}(1-\frac{\partial_z\sqrt{g_{\varphi\varphi}}}
{\sqrt{g_{zz}}}).
\end{equation}
The sense of the metric coefficients $g_{\mu\nu}$ will be explained below.\\\\
The wave function $\psi$, the so-called
bispinor, is composed of two parts:
\begin{equation}\psi=\left(\begin{array}{c}\psi_A \\ \psi_B\end{array}\right),\end{equation}
each corresponding to different sublattices of the curved graphene
sheet. The gauge field $a_{\mu}$ arises from spin
rotation invariance for atoms of different sublattices $A$ and $B$
in the Brillouin zone \cite{8} and the gauge field $a_{\mu}^W$ is connected with the chiral vector $(n,m)$ \cite{9,ando1}:
\begin{equation}a_{\varphi}=N/4,\hspace{1cm}a_{\varphi}^W=-\frac{1}{3}(2m+n).\end{equation}\\\\
The metric $g_{\mu\nu}$ of the 2D surface comes from following
parametrisation, with the help of two parameters $z$, $\varphi$:
\begin{equation}\label{8}(z,\varphi)\rightarrow \overrightarrow{R}=(x(z,\varphi),y(z,\varphi),
z),\end{equation} where
\begin{equation}0<z<\infty,\hspace{1cm}0\leq\varphi<2\pi.\end{equation}
The 4 components of the metric are defined as:
\begin{equation}g_{\mu\nu}=\partial_{\mu}\overrightarrow{R}\partial_{\nu}\overrightarrow{R}.
\end{equation}
For the rotationally symmetric case which will be researched, the non-diagonal components of the metric are
\begin{equation}g_{z\varphi}=g_{\varphi z}=0.\end{equation}\\\\
If we write the wave function in the form
\begin{equation}\label{wavef}\left(\begin{array}{c}\psi_A \\ \psi_B\end{array}\right)
=\frac{1}{\sqrt[4]{g_{\varphi\varphi}}}\left(\begin{array}{c}u(z)e^{i\varphi
j}\\ v(z)e^{i\varphi(j+1)}\\\end{array}\right),\hspace{1cm} j=0,\pm
1,...\end{equation} and substituting (\ref{wavef}) into (\ref{DirC}) we obtain
\begin{equation}\frac{\partial_zu}{\sqrt{g_{zz}}}-\frac{\widetilde{j}}
{\sqrt{g_{\varphi\varphi}}}u
=Ev,\hspace{1cm}-\frac{\partial_zv}{\sqrt{g_{zz}}}-\frac{\widetilde{j}}
{\sqrt{g_{\varphi\varphi}}}v
=Eu,\end{equation}
where
\begin{equation}\widetilde{j}=j+1/2-a_{\varphi}-a_{\varphi}^W.\end{equation}\\\\
Each of the solutions $u,v$ consists of two linearly independent components such that
\begin{equation}u(E,z)=C_1(E)u_1(E,z)+C_2(E)u_2(E,z),\end{equation}
\begin{equation}v(E,z)=\frac{C_1}{E}\left(\frac{\partial_zu_1}{\sqrt{g_{zz}}}-\frac{\widetilde{j}u_1}
{\sqrt{g_{\varphi\varphi}}}\right)+\frac{C_2}{E}\left(\frac{\partial_zu_2}{\sqrt{g_{zz}}}-\frac{\widetilde{j}u_2}
{\sqrt{g_{\varphi\varphi}}}\right),\end{equation}
where for a concrete value of $E$, the functions $C_1(E),C_2(E)$ stand for satisfying the normalization condition
\begin{equation}2\pi\int\limits_{-z_{max}}^{z_{max}}(|u(E,z)|^2+|v(E,z)|^2){\rm d}z=1.\end{equation}\\\\
For a given $z_0$, the $LDoS$ is defined as
\begin{equation}LDoS(E)=|u(E,z_0)|^2+|v(E,z_0)|^2.\end{equation}\\\\

\section{Carbon nanotubes and nanostructured cylinders}\

Nanostructured cylinders consist of rolled-up graphene sheet from which the nanotubes arise by adding cups to the open edges. The properties of carbon nanotubes are presented in \cite{10}. They can be characterized by the chiral vector $\overrightarrow{C}$, defined as
\begin{equation}\overrightarrow{C}=n\overrightarrow{a_1}+m\overrightarrow{a_2}=(n,m),\end{equation}
where $n,m\in\mathcal{N}$ and $\overrightarrow{a_1},\overrightarrow{a_2}$ are the unit vectors of the chosen coordinate system of the original graphene plane \cite{10}. The direction and length of $\overrightarrow{C}$ coincide with the circumference of the nanotube.\\

\begin{figure}
{\includegraphics{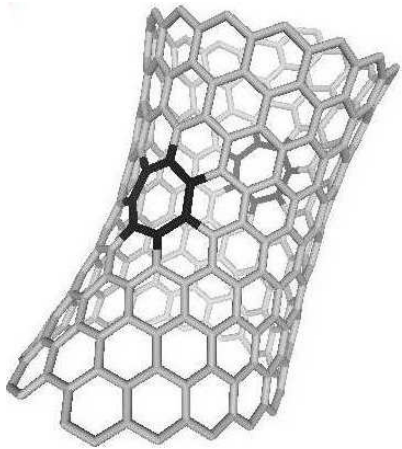}}\caption{Nanocylinder with a small perturbation}\label{fg1}
\end{figure}

According to the symmetrical properties, we distinct 3 kinds of the carbon nanotubes:\\\\
$\bullet$ armchair ($ac$): $\overrightarrow{C}=(n,n)$,\\\\
$\bullet$ zig-zag ($zz$): $\overrightarrow{C}=(n,0)$,\\\\
$\bullet$ chiral: otherwise.\\

One can easily see that $ac$ and $zz$ are the only variants with the mirror symmetry.\\

Furthermore, the carbon nanotubes can be divided into metals or semimetals depending on whether or not $n-m$ is a multiple of $3$. Evidently, all kinds of $ac$ nanotubes are metals.

In this paper, we are concerned with the case of cylinder without and with a small perturbation, caused by 2 heptagonal defects (see Fig. \ref{fg1}).
\\\\

\section{Local density of states}\

The following calculations of $LDoS$ suppose long cylinders with large diameters. The $z$ axis is identical with the axis of the cylinder and the values of $z$ coordinate go from $-z_{max}$ to $z_{max}$. It means that for $z=0$, the results must be similar to the solution for the simple graphene and for $z=\pm z_{max}$, the results should correspond to the edge states as stated in \cite{11}. But as we will see, our method of calculation is not accurate enough to distinct the edge states for the $ac$ and the $zz$ variants, respectively. The reason is that the Dirac equation involves information about the geometrical structure but not about the space coordinates of the individual atoms.

 The calculations will be first done for the defect-free cylinder and then for the perturbed cylinder.\\\\

\subsection{Defect-free cylinder}\

The defect-free cylinder can be described with the help of the parametrization
\begin{equation}\overrightarrow{R}(z,\varphi)=(a\cos\varphi,a\sin\varphi,z),\end{equation}
where $a$ is the radius. The metric coefficients will be then
\begin{equation}g_{zz}=1,\hspace{1cm}g_{\varphi\varphi}=a^2\end{equation}
and
\begin{equation}{\rm det}\,g_{\mu\nu}=g=a^2.\end{equation}
Because the metric coefficients are constant, i.e. $\frac{{\rm\partial}g_{\mu\nu}}{{\rm\partial}x^{\lambda}}=0$, it follows for the coefficients of the Levi-Civita connection and the parameter $\omega$ related to the spin connection
\begin{equation}
\Gamma^{\rho}_{\mu\nu}=0,\hspace{1cm}\omega=\frac{1}{2}.
\end{equation}
From this follows
\begin{equation}\frac{\partial_z u}{\sqrt{g_{zz}}}-\frac{\widetilde{j}}{\sqrt{g_{\varphi\varphi}}}u=Ev,\hspace{1cm}
-\frac{\partial_z v}{\sqrt{g_{zz}}}-\frac{\widetilde{j}}{\sqrt{g_{\varphi\varphi}}}v=Eu,\end{equation}
so
\begin{equation}\partial_z u-\frac{\widetilde{j}}{a}u=Ev,\hspace{1cm}-\partial_z v-\frac{\widetilde{j}}{a}v=Eu,\end{equation}
where
\begin{equation}\widetilde{j}=j + 0.5 + \frac{1}{3} (2 m + n).\end{equation}
The solution has the form
\begin{equation}\label{er1}u(z)=C_1\exp(\alpha z)+C_2\exp(-\alpha z),\end{equation}
\begin{equation}\label{er2}v(z)=\frac{C_1}{E}\left(\alpha-\frac{\widetilde{j}}{a}\right)\exp(\alpha z)-\frac{C_2}{E}\left(\alpha+\frac{\widetilde{j}}{a}\right)\exp(-\alpha z),\end{equation}
where
\begin{equation}\alpha=\alpha(E)=\sqrt{\frac{\widetilde{j}^2}{a^2}-E^2}\end{equation}
The procedure of the calculation of the normalization constants $C_1(E), C_2(E)$ is described in the Appendix A. An interesting task is the case of the zero modes which corresponds to the states of electrons at the Fermi level. They are derived in the Appendix B.

\begin{figure}
{\includegraphics{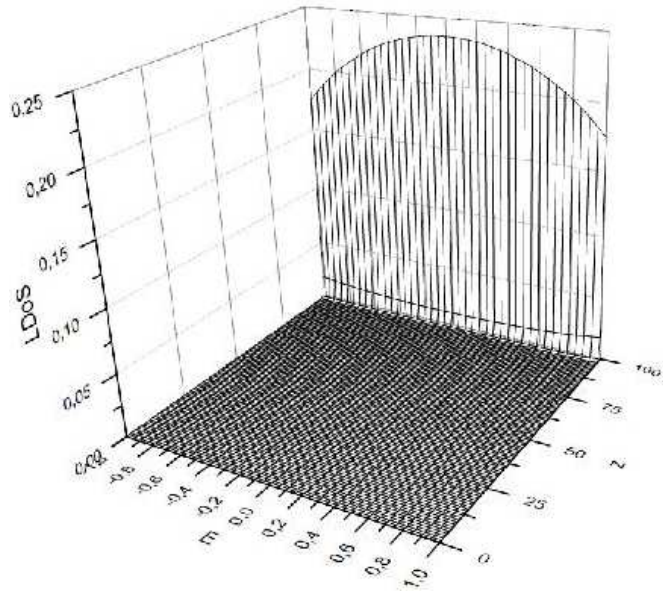}}\caption{$LDoS$
as a function of $E\in (-1,1)$ and $z\in (0,100)$ for defect-free cylinder}\label{fg3}
\end{figure}

In Fig. \ref{fg2} the $LDoS$ is shown as a function of $E$ in the interval $(-1,1)$. In Fig. \ref{fg3} $LDoS$ is plotted as a function of 2 variables, $E$ and $z$. It is case of the armchair configuration. As aforementioned, long cylinders with large diameters are supposed, so we choose high values, i. e. $m=20, n=20, a=14, z_{max}=100$ and $j=0$. Because of the exponential character of $LDoS$, its values are negligible for a long range of $z$ values, so we choose a high $z$ in the $2D$ plot of Fig. \ref{fg2}, i. e. $z=98$. For the same reason, similar values will be chosen for the perturbed case.

\begin{figure}
{\includegraphics{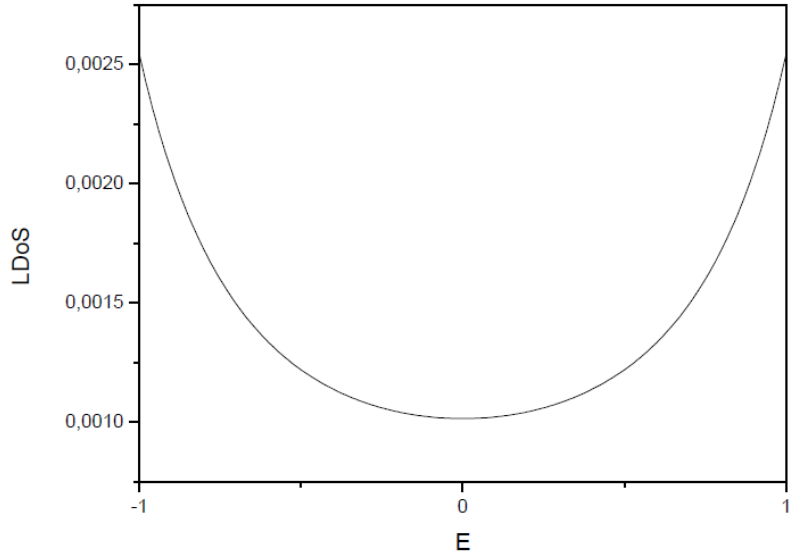}}\caption{$LDoS$
as a function of $E\in (-1,1)$ and $z=98$ for defect-free cylinder}\label{fg2}
\end{figure}

\subsection{Perturbed cylinder}\

In the case of a small perturbation, we have
\begin{equation}\overrightarrow{R}(z,\varphi)=\left(a\sqrt{1+\triangle z^2}\cos\varphi,
a\sqrt{1+\triangle z^2}\sin\varphi,z\right),
\end{equation}
where $\triangle$ is a positive real parameter, $\triangle<<1$. For $\triangle=0$, we get the defect-free cylinder discussed in previous chapter.

Then
\begin{equation}g_{zz}=1+\frac{a^2\triangle^2 z^2}{1+\triangle z^2}\sim 1+a^2\triangle^2 z^2,\hspace{1cm}
g_{\varphi\varphi}=a^2(1+\triangle z^2)
\end{equation}
and
\begin{equation}g=g_{\varphi\varphi}=a^2(1+\triangle z^2).\end{equation}
The only nonzero coefficients of the Levi-Civita connection will be
\begin{equation}\Gamma^z_{zz}\sim a^2\triangle^2z,\hspace{1cm}\Gamma^z_{\varphi\varphi}\sim -a^2\triangle z,\hspace{1cm}
\Gamma^{\varphi}_{z\varphi}=\Gamma^{\varphi}_{\varphi z}\sim a^4\triangle z\end{equation}
and the $\omega$ parameter
\begin{equation}\omega\sim\frac{1}{2}(1-a\triangle^2z).\end{equation}
Then we solve the system of equations
\begin{equation}\label{pert1st}\frac{\partial_zu}{\sqrt{1+a^2\triangle^2z^2}}-
\frac{\widetilde{j}}{a}\frac{u}{\sqrt{1+\triangle z^2}}=Ev,\hspace{1cm}
-\frac{\partial_zv}{\sqrt{1+a^2\triangle^2z^2}}-
\frac{\widetilde{j}}{a}\frac{v}{\sqrt{1+\triangle z^2}}=Eu,\end{equation}
where
\begin{equation}\widetilde{j}=j + 1/2 + \frac{1}{3}(2m + n)-\frac{N}{4},\end{equation}
i.e. we consider the perturbation to be caused by $N$ heptagonal defects.
For small $\triangle$ and neglecting the second order of $\triangle$, it can be simplified as
\begin{equation}\label{er4}\partial_zu-
\frac{\widetilde{j}}{a}\left(1-\frac{1}{2}\triangle z^2\right)u=Ev,\hspace{1cm}
-\partial_zv-\frac{\widetilde{j}}{a}\left(1-\frac{1}{2}\triangle z^2\right)v=Eu.\end{equation}
 The solution is
\begin{equation}u(z)=C_{\triangle 1}D_{\nu_1}(\xi(z))+C_{\triangle 2}D_{\nu_2}(i\xi(z)),\end{equation}
\begin{equation}v(z)=\frac{C_{\triangle 1}}{E}\left(\partial_zD_{\nu_1}(\xi(z))-\frac{\widetilde{j}D_{\nu_1}(\xi(z))}{a}
(1-\frac{1}{2}\triangle^2z^2)\right)+
\frac{C_{\triangle 2}}{E}\left(\partial_zD_{\nu_2}(i\xi(z))-\frac{\widetilde{j}D_{\nu_2}(i\xi(z))}{a}
(1-\frac{1}{2}\triangle^2z^2)\right),\end{equation}
where
\begin{equation}\nu_1=i\frac{a^2\triangle-4 a^2 E^2+4ia\sqrt{\triangle}\widetilde{j}+4\widetilde{j}^2}{8a\sqrt{\triangle} \widetilde{j}},
\hspace{1cm}\nu_2=-i\frac{a^2\triangle-4 a^2 E^2-4ia\sqrt{\triangle}\widetilde{j}+4\widetilde{j}^2}{8a\sqrt{\triangle} \widetilde{j}},
\end{equation}
\begin{equation}\xi(z)= (-\triangle)^{1/4}\left(\sqrt{\frac{a}{
 2\widetilde{j}}} + \sqrt{\frac{2\widetilde{j}}{a}}z\right),
\end{equation}
$D_{\nu}(\xi)$ being the parabolic cylinder function \cite{12}. The functions $C_{\triangle 1}=C_{\triangle 1}(E),\,C_{\triangle 2}=C_{\triangle 2}(E)$ will be calculated in the same way as for the defect-free cylinder, i. e. from the normalization condition \begin{equation}\int\limits_{-z_{max}}^{z_{max}}(|u(E,z)|^2+|v(E,z)|^2){\rm d}z{\rm d}\varphi=
4\pi\int\limits_0^{z_{max}}(|u(E,z)|^2+|v(E,z)|^2){\rm d}z=1.\end{equation}

\begin{figure}
{\includegraphics{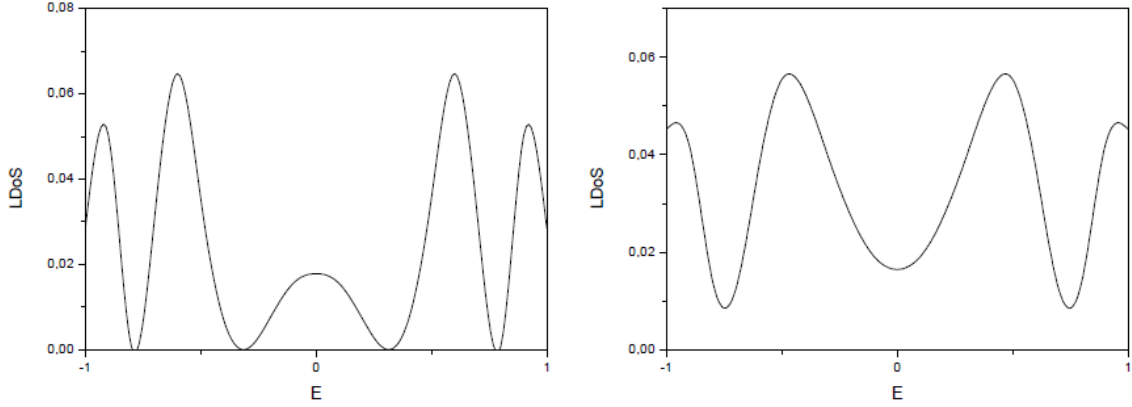}}\caption{$LDoS$
as a function of $E\in (-1,1)$ and $z=98$ for perturbed cylinder with $\triangle=0.05$ (left) and $\triangle=0.1$ (right)}\label{fg4}
\end{figure}

\begin{figure}
{\includegraphics{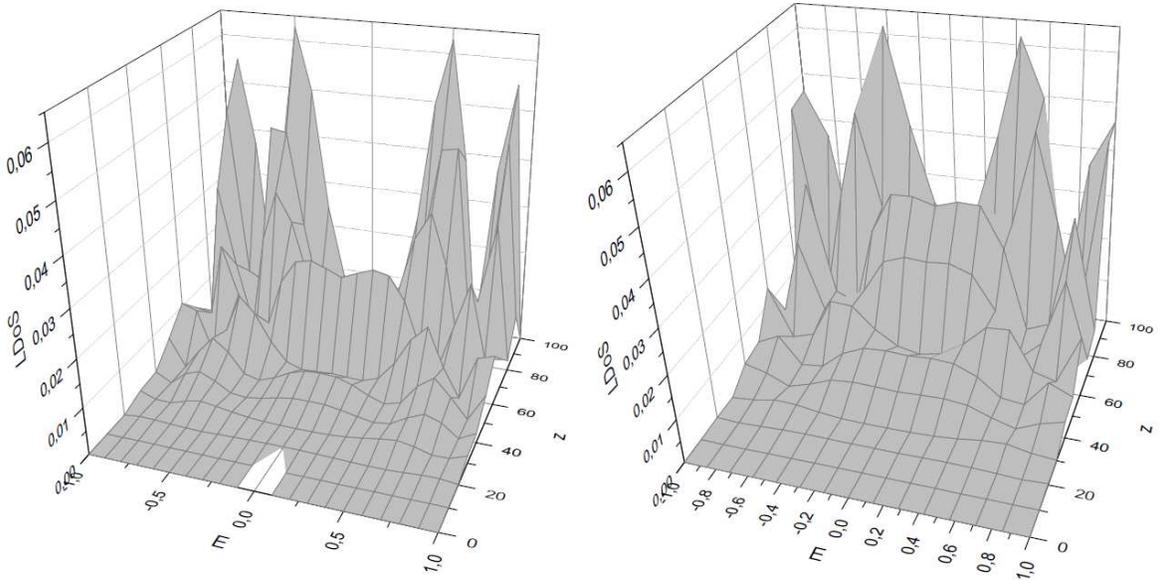}}\caption{$LDoS$
as a function of $E\in (-1,1)$ and $z\in (0,100)$ for perturbed cylinder with $\triangle=0.05$ (left) and $\triangle=0.1$ (right)}\label{fg5}
\end{figure}

In Fig. \ref{fg4}, a comparison of the $2D$ plots of $LDoS$ is made for different values of $\triangle$ and similarly in Fig. \ref{fg5}, where the $3D$ plots are compared. The number of defects $N=2$ and the other values are same as in the defect-free case. On the whole, we can conclude from the comparison that $LDoS$ is slightly increasing for higher $\triangle$. Although this is only the first order approximation, we will see that the real solution of (\ref{pert1st}) behaves in a similar manner. The case of zero modes appropriate to the states at the Fermi level is calculated in the Appendix C.\\

Comparing the calculated expressions (\ref{DF0}), (\ref{P0}) for zero modes corresponding to the defect-free and perturbed case, respectively, we see that it has a character of an exponential function related to a polynomial in $z$ its order depends on the chosen precision. This means that the simplest case of the first order polynomial corresponds to the defect-free case, the higher orders of the polynomial correspond to the perturbed case. For higher $z$, the function $u_0(z)$ corresponding to the sublattice $A$ is increasing, whereas the function $v_0(z)$ corresponding to the sublattice $B$ is decreasing.\\\\

\section{Case of nanoribbons}\

For large diameters, the presented results for the defect-free cylinder should correspond with the density of states for nanoribbons (\textit{ac} or \textit{zz}). The reason is that because of a very small curvature, a nanocylinder with a large diameter we can take as an approximation of an infinitely long nanoribbon (see e.g. \cite{nature}).

Let us give a small review about the structure and properties of the nanoribbons. Graphene nanoribbons are materials, theirs molecules have the form of thin, rectangular pieces cut from the graphene layer, so their structure is analogical. Similarly as in the case of graphene nanotubes, \textit{ac} and \textit{zz} forms are distinguished, theirs difference consists in the form of the edges. Denoting $N-1$ the number of hexagonal unit cells by which the nanoribbon width is constituted, metallic and semiconducting forms can be marked  again: a concrete form of nanoribbon is metallic, if $N+1$ is a multiple of $3$, so this condition differs from the condition mentioned in the third chapter for carbon nanotubes.

To calculate the energy spectrum, the tight-binding approximation is used. Deriving the form of Hamiltonian for the given variant (\textit{ac} or \textit{zz}), we get the set of equations of motion. For \textit{ac}, it has the form \cite{13}
\begin{equation}\label{sys}E\psi_{m,A}=-e^{-ik/2}\psi_{m,B}-\psi_{m-1,B}-\psi_{m+1,B},\hspace{1cm}
E\psi_{m,B}=-e^{ik/2}\psi_{m,A}-\psi_{m-1,A}-\psi_{m+1,A},\end{equation}
where $m=1,\ldots,N$ denotes the position of the corresponding $A$ or $B$ atom in the nanoribbon, $\psi_{m,A}$ or $\psi_{m,B}$, resp. is its wave function and $k=\frac{2\pi}{L_y}n,\hspace{5mm}n=0,\pm 1,\ldots,\pm (\frac{L_y}{2}-1),\frac{L_y}{2}$ ($L_y$ is the number of the hexagonal unit cells in the longitudinal direction), is the longitudinal wave number. With the help of the boundary conditions and respecting the fact that the determinant of the system (\ref{sys}) should be equal to zero, we get that the energy spectrum contains the values
\begin{equation}\label{spectAC}E=\pm\sqrt{1+2\varepsilon_r\cos\left(\frac{k}{2}\right)+\varepsilon_r^2},\end{equation}
where $\varepsilon_r=2\cos\frac{r}{N+1}\pi,\hspace{5mm}r=1,2,\ldots N$. Using this procedure for \textit{zz}, the analogous energy spectrum appears, but the formulas are a little bit modificated. In \cite{ando2}, similar results are obtained for the carbon nanotubes.\\

\begin{figure}
{\includegraphics{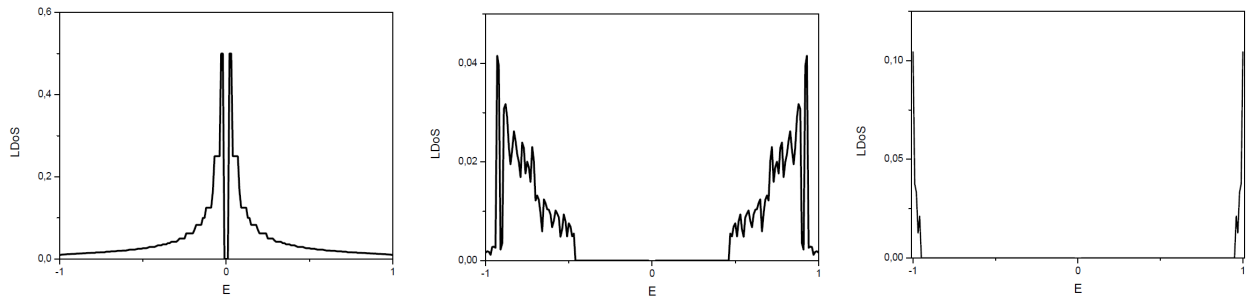}}\caption{$LDoS$
as a function of $E\in (-1,1)$ for $k=0$ (left), $k=\frac{\pi}{2}$ (middle) and $k=\pi$ (right) for \textit{ac} nanoribbons}\label{fg7}
\end{figure}

Now the $LDoS$ for \textit{ac} nanoribbons will be calculated and the results will be compared with the above stated calculations for the defect-free nanocylinder. The corresponding formulae is
\begin{equation}\label{LDoS}LDoS(E,k)=\frac{\delta(E-E(k))}{D(E)},\end{equation}
where
\begin{equation}D(E)=\lim_{\eta\rightarrow0}2{\rm Im}\int_{-\pi}^{\pi}{\rm d}k\frac{k}{E-E(k)-i\eta},\end{equation}
$\delta(E-E(k))$ is the Dirac delta function and $k$ is the wavenumber. Let us stress that the integration over $k$ is used here instead of the integration over $z$. Both kinds of integration are related by the Fourier transform, as follows from \cite{mele}. For different $k$, we see the plots of $LDoS$ in Fig. \ref{fg7}. Evidently, excluding the case $k=0$, the results more or less correspond to the solution given by the formulas (\ref{er1}), (\ref{er2}).\\

In \cite{13}, the possible solutions for $zz$ are divided into the cases of the extended and the localized states. The form of the corresponding wavefunction for the localized states is
\begin{equation}\left(\begin{array}{c}\psi_{m,A} \\ \psi_{m,B}\end{array}\right)
=C\left(\begin{array}{c}\sinh \frac{\pi r(N+1-m)}{N+1}\\ \sinh \frac{\pi rm}{N+1}\\\end{array}\right).\end{equation}
Since for large $z$ we can write
\begin{equation}\sinh z \sim \exp z,\end{equation}
this form corresponds to the solution for the defect-free case (\ref{er1}), (\ref{er2}). So, we see an important coincidence with the results found for nanoribbons in \cite{13}. The results presented in this paper are related to the case of $ac$ configuration, but as stressed in the beginning of the fourth section, our method of calculation does not distinguish between $ac$ and $zz$ and the plots for case of $zz$ would be similar.\\

Because of the discrete spectrum of carbon nanoribbons and from the fact that it can be sometimes understood as an approximation of the nanocylinders a question arises about a possibility of calculation the low energy electronic states for the case of small defects. Similar task was done in \cite{15} for the case of the spheroidal geometry. This could be possible in case of using different methods for calculation of the $LDoS$. In this paper, the resulting energy spectrum can be understood as continuous, so there is not any possibility of shifting the energy levels due to the defects.\\\\

\section{Conclusion}\

Comparing the plots of $LDoS$ in Fig. \ref{fg7} with the plots in Fig. \ref{fg2}, we see that both Figures are similar but not completely same. The reason is in difference between the used procedures: for the purpose of Fig. \ref{fg7}, the energy spectrum is first calculated in (\ref{spectAC}) and then the corresponding $LDoS$ is found according to (\ref{LDoS}), but the procedure used in chapter IV is opposite, i.e. first the $LDoS$ is calculated and then the energy spectrum is estimated from the plots (as in \cite{mele}). Because such a procedure is only a rough approximation, it does not enable us to find the localized states for zero energy as in (\ref{LDoS}). So, only the method described in the previous chapter enables us to distinguish metals and semimetals. Furthermore, according to \cite{13}, the condition for nanoribbons to be metal is different from the condition for nanotubes, so the results could never be exactly same.\\

\begin{figure}
{\includegraphics{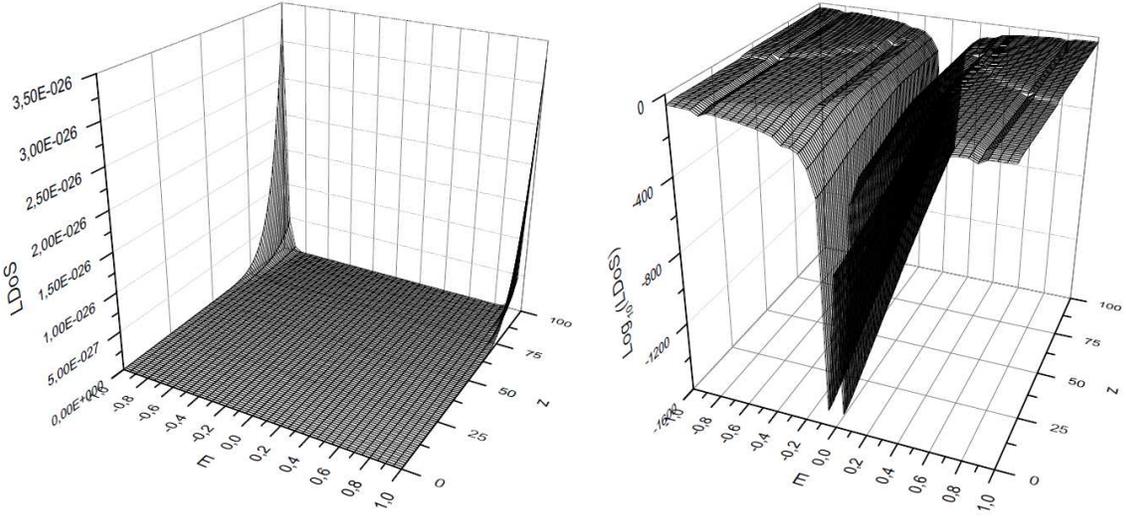}}\caption{$LDoS$
as a function of $E\in (-1,1)$ and $z\in (0,100)$ for perturbed cylinder with $\triangle=0.05$ on linear (left) and logarithmic (right) scale (numerical solution)} \label{fg8}
\end{figure}

\begin{figure}
{\includegraphics{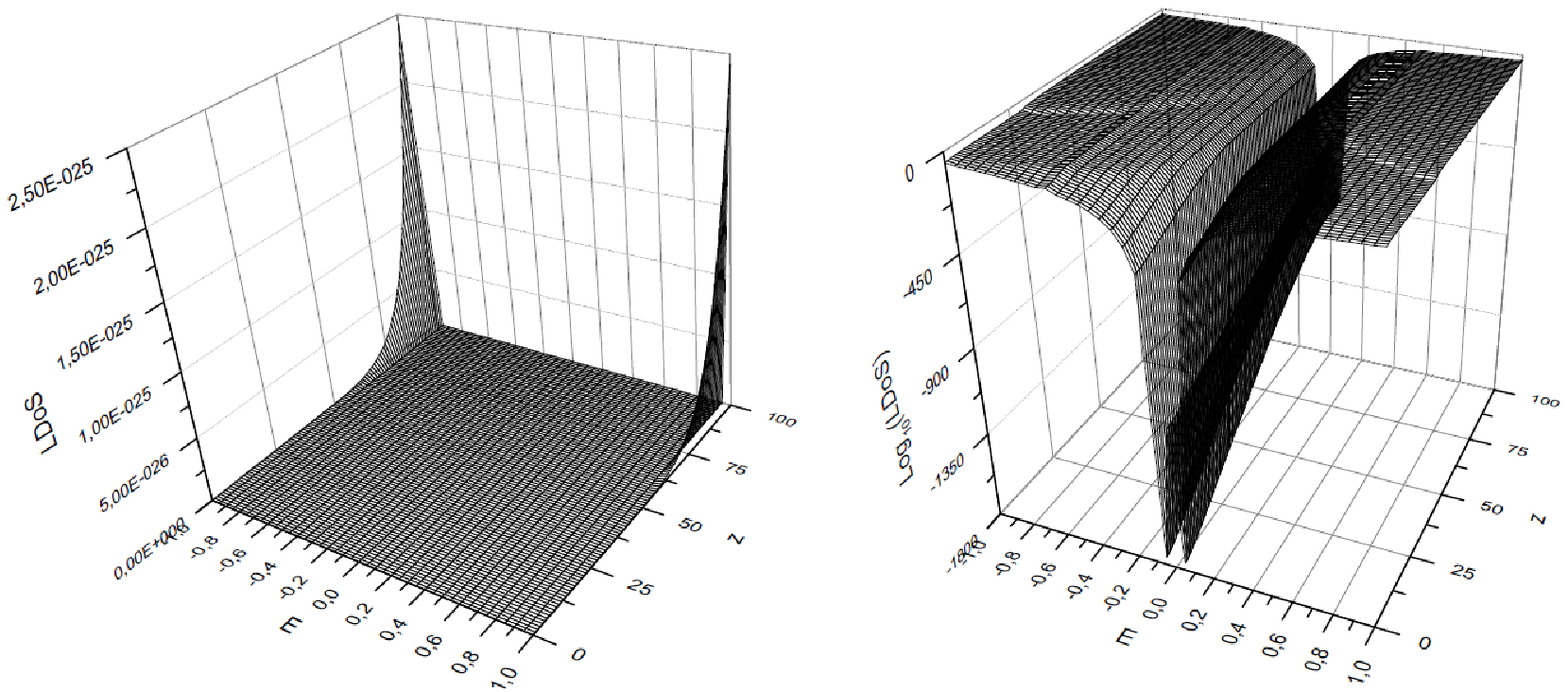}}\caption{$LDoS$
as a function of $E\in (-1,1)$ and $z\in (0,100)$ for perturbed cylinder with $\triangle=0.1$ on linear (left) and logarithmic (right) scale (numerical solution)} \label{fg9}
\end{figure}

The presented results for perturbed cylinder are only the first order approximation of the real solutions which are presented in Figs. \ref{fg8} and \ref{fg9}. Because of the complicated form of (\ref{pert1st}), the solutions were acquired in a numerical way. The analytical solutions we can get in a more precise form, if we increase the order of the parameter $\triangle$ in (\ref{er4}). It follows from the plots on  logarithmic scale in Figs. \ref{fg8}, \ref{fg9} (and also from Fig. \ref{fg4} for the first order solution) that for $E=0$, a localized state with a small amplitude appears. From this we see the evidence that the metallic properties of the nanocylinder are increasing by adding a perturbation into the structure.\\

To conclude, the $LDoS$ was researched for nanocylinders without and with a defect. We found that for the defect-free case, the electronic properties can sometimes correspond to the case of infinitely long nanoribbons. For the case of perturbation, the metallic properties are more manifested.\\

The generation of the nanocylinders is in a close correspondence with the generation of the nanotubes. The most exploited method is the chemical vapor deposition (CVD) which is based on the production of the required material from the surface of a substrate by volatile precursors. Usually it is realized as a thermal reaction called catalytic growth process, when the growth of carbon nanotubes is catalyzed in a heated flow furnace by a carbon surface diffusion \cite{chiang,muram}. This process is activated by iron and nickel nanoparticles with the help of acetylene and hydrogen as the precursors. The nanocylinders are produced in this process together with the nanotubes or they arise from the nanotubes with the help of other thermal or chemical processes. The surface of the nanotubes is usually unreactive, but the change of its structure can be provoked by catalytically active atoms of rhenium inserted into the nanotubes \cite{chamber}. For the case of the perturbed nanocylinder, as presented in this paper, its structure can be determined by the presence of two heptagonal defects on the surface, but this structure sometimes appears without defects and it is only given by different lengths of the bonds between the atoms. The calculations performed in this paper are valid anyway.

\appendix

\section{Normalization constants of defect-free cylinder}\

We will suppose that for the solution holds
\begin{equation}u(z)=u(-z),\hspace{1cm}v(z)=v(-z).\end{equation}
Then the normalization condition gives
\[\int\limits_{-z_{max}}^{z_{max}}(|u|^2+|v|^2){\rm d}z{\rm d}\varphi=
2\int\limits_0^{z_{max}}(|u|^2+|v|^2)2\pi{\rm d}z=\]
\[=\frac{8\pi\widetilde{j}}{aE^2}\int\limits_0^{z_{max}}\left[C_1^2(E)(\frac{\widetilde{j}}{a}-\alpha(E))\exp(2\alpha(E) z)+
C_2^2(E)(\frac{\widetilde{j}}{a}+\alpha(E))\exp(-2\alpha(E) z)+2C_1(E)C_2(E)\frac{aE^2}{\widetilde{j}}\right]{\rm d}z=\]
\[=\frac{4\pi\widetilde{j}}{a\alpha(E)E^2}\{C_1^2(E)(\frac{\widetilde{j}}{a}-\alpha(E))[\exp(2\alpha(E)z_{max})-1]-
C_2^2(E)(\frac{\widetilde{j}}{a}+\alpha(E))[\exp(-2\alpha(E)z_{max})-1]\}+\]
\begin{equation}\label{er3}+16\pi C_1(E)C_2(E)z_{max}=1.\end{equation}
From (\ref{er1}),(\ref{er2}) and (\ref{er3})
\[LDoS(E,z)=|u(E,z)|^2+|v(E,z)|^2=\frac{1}{8\pi z_{max}}+\]
\[+C_1(E)^2\left(\frac{\widetilde{j}}{a}-\alpha(E)\right)\frac{\widetilde{j}}{aE^2}\left(2\exp(2\alpha(E)z)-
\frac{\exp(2\alpha z_{max})-1}{2\alpha z_{max}}\right)+\]
\begin{equation}+C_2(E)^2\left(\frac{\widetilde{j}}{a}+\alpha(E)\right)\frac{\widetilde{j}}{aE^2}\left(2\exp(-2\alpha(E)z)+
\frac{\exp(-2\alpha z_{max})-1}{2\alpha z_{max}}\right).\end{equation}
Now, if we suppose $C_1(E)\in\mathcal{R}$ for all $E\in (-1,1)$, it follows from (\ref{er3})
\begin{equation}C_1(E)\leq f(E)=\sqrt{\frac{a\alpha(E)E^2}{4\pi\widetilde{j}(\frac{\widetilde{j}}{a}-\alpha(E))(\exp(2\alpha(E) z_{max})-1)}}.\end{equation}
So, for the given choice of the parameters, we find the dependence $f=f(E)$ and we can choose $C_1(E)=min\{f(E),E\in (-1,1)\}$. Usually, we choose $C_1(E)=C_2(E)$, but this choice is not necessary.\\

\section{Zero modes for defect-free case}\

The solution of the zero modes has the form
\begin{equation}\label{DF0}u_0(z)=C_u\exp\left(\frac{\widetilde{j}}{a}z\right),\hspace{1cm}
v_0(z)=C_v\exp\left(-\frac{\widetilde{j}}{a}z\right).\end{equation}
In accordance with the previous procedure,
\begin{equation}4\pi\int\limits_0^{z_{max}}(|u_0(z)|^2+|v_0(z)|^2){\rm d}z=
\frac{2\pi a}{\widetilde{j}}\left\{C_u^2\left[\exp\left(\frac{2\widetilde{j}}{a}z_{max}\right)-1\right]-
C_v^2\left[\exp\left(-\frac{2\widetilde{j}}{a}z_{max}\right)-1\right]\right\}=1\end{equation}
and
\begin{equation}LDoS(z)=C_u^2\left[\exp\left(\frac{2\widetilde{j}}{a}z_{max}\right)-1\right]+\frac{\widetilde{j}/(2\pi a)}
{\exp\left(\frac{2\widetilde{j}}{a}z_{max}\right)-1}.\end{equation}
Then
\begin{equation}C_u\leq\sqrt{\frac{\widetilde{j}/(2\pi a)}
{\exp\left(\frac{2\widetilde{j}}{a}z_{max}\right)-1}}.\end{equation}
\\\\

\section{Zero modes for perturbed case}\

The zero modes for the case of perturbation will be calculated from the equations (\ref{er4}) without the r.h.s, i.e.
\begin{equation}\partial_zu_0-
\frac{\widetilde{j}}{a}\left(1-\frac{1}{2}\triangle z^2\right)u_0=0,\hspace{1cm} -\partial_zv_0-
\frac{\widetilde{j}}{a}\left(1-\frac{1}{2}\triangle z^2\right)v_0=0,\end{equation}
with the solution
\begin{equation}\label{P0}u_0(z)=C_{\triangle u}\exp\left[\frac{\widetilde{j}}{a}z\left(1-\frac{\triangle z^2}{6}\right)\right],\hspace{1cm}
v_0(z)=C_{\triangle v}\exp\left[-\frac{\widetilde{j}}{a}z\left(1-\frac{\triangle z^2}{6}\right)\right].\end{equation}
We require
\[4\pi\int\limits_0^{z_{max}}(|u_0(z)|^2+|v_0(z)|^2){\rm d}z=
4\pi\int\limits_0^{z_{max}}\left\{C_{\triangle u}^2\exp\left[\frac{2\widetilde{j}}{a}z\left(1-\frac{\triangle z^2}{6}\right)\right]+C_{\triangle v}^2\exp\left[-\frac{2\widetilde{j}}{a}z\left(1-\frac{\triangle z^2}{6}\right)\right]\right\}{\rm d}z=\]
\begin{equation}=4\pi\int\limits_0^{z_{max}}\left\{C_{\triangle u}^2\exp\left(\frac{2\widetilde{j}}{a}z\right)
\exp(-\delta z^3)+C_{\triangle v}^2\exp\left(-\frac{2\widetilde{j}}{a}z\right)\exp(\delta z^3)\right\}{\rm d}z=1,\end{equation}
where we denoted $\frac{\widetilde{j}\triangle}{3a}=\delta$.
In case $\delta z_{max}^3<<1$, an approximate expansion can be done
\begin{equation}\exp(\delta z^3)=1+\delta z^3\end{equation}
and the last integral equals
\[4\pi\int\limits_0^{z_{max}}\left\{C_{\triangle u}^2\exp\left(\frac{2\widetilde{j}}{a}z\right)
(1-\delta z^3)+C_{\triangle v}^2\exp\left(-\frac{2\widetilde{j}}{a}z\right)(1+\delta z^3)\right\}{\rm d}z=\]
\[=\frac{a\pi}{2\widetilde{j}^4}C_{\triangle u}^2\left\{\exp\left(\frac{2\widetilde{j}z_{max}}{a}\right)\left[3\delta(a-\widetilde{j}z_{max})
(a^2-a\widetilde{j}z_{max}+\widetilde{j}^2z_{max}^2)+\widetilde{j}^3(4-\delta z_{max}^3)\right]-3a^3\delta-4\widetilde{j}^3\right\}-\]
\begin{equation}-\frac{a\pi}{2\widetilde{j}^4}C_{\triangle v}^2\left\{\exp\left(-\frac{2\widetilde{j}z_{max}}{a}\right)\left[3\delta(a+\widetilde{j}z_{max})
(a^2+a\widetilde{j}z_{max}+\widetilde{j}^2z_{max}^2)+\widetilde{j}^3(4+\delta z_{max}^3)\right]-3a^3\delta-4\widetilde{j}^3\right\}=1.
\end{equation}
If $\delta z_{max}^3<<1$ is not true, we do the expansion
\begin{equation}\exp(\delta z^3)=1+\delta z^3+\frac{(\delta z^3)^2}{2!}+...+\frac{(\delta z^3)^n}{n!},\end{equation}
where $\frac{(\delta z^3)^n}{n!}<<1$, substitute it into the integral and make an approximate calculation. Calculations of $LDoS,\,C_{\triangle u}$ and $C_{\triangle v}$ are then done in the usual way.\\\\

\end{document}